\documentclass[pra,amsmath,twocolumn,showpacs,floatfix]{revtex4}
\pdfoutput=1

\usepackage[utf8]{inputenc}
\usepackage{graphicx}
\usepackage{bm}

\usepackage{hyperref}
\hypersetup{
   colorlinks=true,
   linkcolor=black,
   citecolor=black,
   urlcolor=blue,
   pdfauthor={Janus H. Wesenberg},
   pdftitle={Ideal intersections for radio-frequency trap networks}
}

\graphicspath{%
  {figures/}{figures/converted/}%
  {../20080924_SussexTalk/figures/}{../20080924_SussexTalk/figures/converted/}%
  {../figures/}{../figures/converted/}%
}

\renewcommand{\vec}[1]{\ensuremath{\boldsymbol{#1}}}
\newcommand{\vhat}[1]{\ensuremath{\hat{\vec{#1}}}}

\DeclareMathOperator{\bigo}{\mathcal{O}}

\newcommand{\topp}[1]{^{({#1})}}

\newcommand{\grad}{\vec{\nabla}}

\newcommand{\ve}{\ensuremath{V}}

\newcommand{\erf}{\ensuremath{\vec{E}}}

\newcommand{\nveci}[1]{\ensuremath{\hat{\vec{N}}\topp{#1}}}
\newcommand{\tveci}[1]{\ensuremath{\hat{\vec{T}}\topp{#1}}}
\newcommand{\curvi}[1]{\ensuremath{\kappa\topp{#1}}}

\newcommand{\tvecij}[2]{\ensuremath{\hat{T}\topp{#1}_{#2}}}

\newcommand{\vpp}{\ensuremath{U}}

\newcommand{\icurvi}[1]{\ensuremath{\vec{\gamma}\topp{#1}}}

\newcommand{\strengthx}{\ensuremath{\alpha_{\text{X}}}}
\newcommand{\strengthxx}{\ensuremath{\alpha_{\text{X}}\topp{x}}}
\newcommand{\strengthxy}{\ensuremath{\alpha_{\text{X}}}\topp{y}}
\newcommand{\strengthhomo}{\ensuremath{\alpha_{\text{H}}}}

\newcommand{\basisx}{\ensuremath{\Theta_{\text{X}}}}
\newcommand{\basisxx}{\ensuremath{\Theta_{\text{X}}\topp{x}}}
\newcommand{\basisxy}{\ensuremath{\Theta_{\text{X}}}\topp{y}}

\newcommand{\kerq}{N_q}

\newcommand{\curvidx}{l}

\begin{document}
\title{%
  Ideal intersections for radio-frequency trap networks
}
\date{\today}

\author{J.~H.~Wesenberg}
\email[]{janus.wesenberg@materials.ox.ac.uk}
\affiliation{Department of Materials, University of Oxford, Oxford OX1
  3PH, United Kingdom}
\date{\today}

\begin{abstract}
  We investigate the possible form of ideal intersections for
  two-dimensional rf trap networks suitable for quantum information
  processing with trapped ions.
  We show that the lowest order multipole component of the rf field
  that can contribute to an ideal intersection is a hexapole term
  uniquely determined by the tangents of the intersecting paths.
  The corresponding ponderomotive potential does not provide any
  confinement perpendicular to the paths if these intersect at right
  angles, indicating that ideal right-angle X-intersections are
  impossible to achieve with hexapole fields.
  Based on this result, we propose an implementation of an ideal
  oblique-X intersection using a three-dimensional electrode
  structure.
\end{abstract}


\maketitle

Intersections between network paths are a key ingredient in
two-dimensional (2D) rf trap
networks which have been proposed to allow large scale quantum
information processing (QIP) with trapped ions
\cite{kielpinski02:archit_a_large_scale_trap,leibfried07:transport,ospelkaus08:trapped-ion}.
%
RF traps confine ions by a combination of rf and quasistatic electric
fields \cite{fischer59:die,wuerker59:electrodynamic} and in an ideal
network the rf field and the associated ponderomotive potential vanish
in the trapping zones and on dedicated paths between zones but nowhere
else.
Such ideal trap networks have been demonstrated in one dimension (1D) with segmented
linear rf traps where 
%
%
transport as well as splitting and joining of groups of ions have been
demonstrated to be possible with a very high degree of control
\cite{rowe02:trans_quant_states_separ_ions,barrett04:deter_quant_telep_atomic_qubit,reichle06:transport}.
In contrast, no ideal 2D trap networks have been identified, although
a number of possible intersection geometries for 2D trap networks have been
investigated \cite{%
  chiaverini05:surfac_elect_archit_for_ion,%
  hensinger06:t_junct_trap_array_dimen,%
  pearson06:experimental,%
  hucul07:transport,%
  amini07:multilayer,wesenberg07:analytical}.
The observed shortcomings fall in two broad categories.
For T \cite{hensinger06:t_junct_trap_array_dimen}, Y
\cite{amini07:multilayer,wesenberg07:analytical_theory_multipole} and
some X \cite{chiaverini05:surfac_elect_archit_for_ion,pearson06:experimental},
 intersections, a residual rf field is observed in the paths
through the intersection which can possibly lead to motional heating
\cite{bluemel89:chaos_and_order_laser_cooled}, in addition to
complicating or hindering controlled ion transport
\cite{hucul07:transport}.
In contrast, some high-symmetry X geometries offer truly field-free
paths but fail to be ``fully confining'' in the sense that there are
unwanted lines of zero field allowing ions to escape the trap network
\cite{chiaverini05:surfac_elect_archit_for_ion}.

In this paper we show that the unique simplest form of an ideal
intersection for 2D trap networks is an oblique X and propose an 
implementation of this intersection based on a
three-dimensional (3D) electrode structure that  faithfully implements the ideal
intersection.

The paper is structured as follows:
Sec.~\ref{sec:rf-traps} defines the basic problem of designing
ideal rf trap intersections.
In Sec.~\ref{sec:uniq-hexap-inters} we constructively identify a
hexapole term uniquely determined by the intersection angle as the
only multipole term of hexapole or lower order which can contribute to
a zero-field intersection.
In Sec.~\ref{sec:discussion} we investigate the properties of the
identified hexapole intersection and show that fully confining
hexapole intersections are only possible at oblique intersection
angles. Lastly, in Sec.~\ref{sec:impl-with-3d} we describe an
implementation of the ideal intersection based on a 3D electrode
structure.

\section{RF trap networks}
\label{sec:rf-traps}

Although rf traps use a combination of quasistatic and rf electrical
fields to confine charged particles, we will here only be concerned
with the rf component.
For typical QIP ion traps, the dimensions ($<\text{mm}$) and rf
frequencies ($>\text{MHz}$) are such that we can adequately treat the
rf field as quasistatic and express the time dependent rf field as
$\cos(\Omega t)\, \erf(\vec{r})$, where $\Omega$ and $\erf$ denote the
rf frequency and spatially varying amplitude, respectively.
The rf period $2\pi/\Omega$ is made to be the fastest time scale of
the trap, so that in the adiabatic limit the effect of the field can
be described by an external ponderomotive potential $\vpp$ given by
\begin{equation}
  \label{eq:ppdef}
  \vpp(\vec{r}) \equiv \frac{Q^2}{4 M \Omega^2}\,\left\lvert
    \erf(\vec{r})
  \right\rvert^2,
\end{equation}
where $M$ and $Q$ denote the mass and charge of the trapped particle
\cite{dehmelt67:dehmel,ghosh95:ion}. For the purpose of this paper, it
suffices to note that $\vpp(\vec{r})$ is proportional to the square of
the local rf field amplitude.
Since confinement by electrostatic fields is impossible, the rf field
must always be present to allow trapping, and we will assume
$\erf(\vec{r})$ to be constant as is usually the case in QIP ion
traps. In this case, $\vpp$ constitutes a constant ``landscape'' in
which ions can be shuffled around by manipulating a quasistatic
control field provided by dedicated trap electrodes. As mentioned
above, we are here only interested in the structure of this landscape,
which is determined entirely by the electrode geometry and is
independent of the amplitude and frequency of the applied rf field.

For an ideal trap network, as introduced in the Introduction, we
require that the $\erf(\vec{r})$ vanishes for $\vec{r}$ on a network path
and only there.
Ideal networks have a number of qualities.  Firstly, the ions do not
experience any rf-induced micromotion. Micromotion is especially
critical in trap regions where gate operations are performed
\cite{wineland98:exper_issues_coher_quant_state}, but could also
potentially lead to heating effects
\cite{bluemel89:chaos_and_order_laser_cooled}.
Secondly, the ponderomotive potential associated with a nonzero rf field
along trap paths complicates fully controlled ion transport because the
applied control fields must be engineered to compensate for any
curvature of $\vpp$ along the path in order to avoid motional heating
of the ions \cite{reichle06:transport,hucul07:transport}.  Note that
such compensation is impossible for the transport of multispecies
crystals, which according to Eq.~(\ref{eq:ppdef}) experience different
ponderomotive potentials due to mass differences.
Lastly, the requirement that $\vpp(\vec{r})$ vanishes only on
the network paths ensures that ions are confined, even in the absence
of control fields.

An ideal 1D trap network can be implemented with segmented linear rf
traps \cite{rowe02:trans_quant_states_separ_ions} where the rf field
forms a quadrupole with a line of zero rf field along the trap axis
\cite{paul90:electromagnetic}, so that the isosurfaces of $\vpp$ are
concentric cylinders around the axis.  In this case, transport as well
as splitting and joining of groups of ions have been demonstrated to
be possible with a very high degree of control
\cite{rowe02:trans_quant_states_separ_ions,barrett04:deter_quant_telep_atomic_qubit,reichle06:transport}.

We should note that although our quest in the following is for ideal
trap networks, none of the conditions for a network to be ideal in the
sense introduced above are absolute requirements for a network to be
suitable for even large scale QIP.
Firstly, it has been proposed that transport in a network that is not
fully confining can be achieved by ``surfing'' around any leaks
\cite{hucul07:transport}.
Secondly, even in the presence of residual a rf field along the network
path, it can be possible to perform highly controlled ion transport so
that no or minimal heating takes place
\cite{reichle06:transport,hucul07:transport}.
Nevertheless, it seems that an ideal network would have a number of
operational benefits.

\section{Unique hexapole intersection}
\label{sec:uniq-hexap-inters}

As a first step to constructing an ideal 2D intersection, we will 
study the possible local structure of $\erf(\vec{r})$ at the
intersection of two smooth network paths in terms of a multipole
expansion.
Our goal is to find the lowest order multipole term that can
contribute to $\erf(\vec{r})$. Identifying the lowest possible
multipole order is desirable for two reasons.
Firstly, the confinement that can be produced for a given
ion-electrode distance is stronger for lower-order multipole fields
\cite{wesenberg07:analytical_theory_multipole}. Keeping the
ion-electrode distance as large as possible is important since ion
heating has been repeatedly demonstrated to increase dramatically with
decreasing ion-electrode distance
\cite{turchette00:heatin_trapp_ions_from_quant,deslauriers06:scalin_suppr_anomal_quant_decoh,epstein07:simpl_ion_heatin_rate_measur}.
Secondly, the construction of an intersection where all low-order
multipole terms vanish is technically challenging. By trapping with
the lowest order multipole possible, a smaller number of terms must be
made to vanish. 

With this in mind, we expand $\erf(\vec{r})$ around the intersection
point chosen as the origin, 
\begin{equation}
  \label{eq:erfexp}
  E_i = d_i + q_{i,j} r_j + \tfrac{1}{2} h_{i,j,k} r_j r_k + \bigo(\vec{r}^3),
\end{equation}
where the tensors $q_{i,j} = \partial_j E_i$ and
$h_{i,j,k}=\partial_j \partial_k E_i$ describe the quadrupole and
hexapole components of the field.
Here, as in the remainder of the paper, we have adopted the convention
that summation over repeated indices is implied, 
e.g.~$q_{i,j}r_j= q_{i,1}x +q_{i,2} y +q_{i,3} z$.

It follows from Maxwell's equations that $\erf(\vec{r})$ must be
irrotational and divergence-free in the trap region,
i.e., $\vec{\nabla} \times \erf(\vec{r}) =\vec{0}$ and $\vec{\nabla}
\cdot \erf(\vec{r}) =\vec{0}$.
These requirements give rise to constraints on the quadrupole and
hexapole tensors.
To find these constraints, we note that the requirements are
equivalent to requiring that $\erf(\vec{r})=-\grad \ve(\vec{r})$ for
an electric potential $\ve$ fulfilling the Laplace condition
$\partial_i \partial_i \ve(\vec{r}) = 0$.
For the quadrupole terms, this implies that $q_{i,j}$ must be symmetric and
traceless,
\begin{equation}
  \label{eq:quadconstr}
  q_{i,j}=q_{j,i} \quad \text{ and } \quad  q_{i,i}=0.
\end{equation}
This leaves five free parameters as expected since a basis of the
quadrupole field components can be formed from the five second-order spherical harmonics.
For the hexapole terms, we find similarly that $h_{i,j,k}$ must be
symmetric, and that the partial traces must vanish since
$\partial_i \partial_j \partial_j \ve(\vec{r}) = \partial_i \nabla^2
\ve(\vec{r})$,
\begin{equation}
  \label{eq:hexconstr}
  h_{i,j,k}=h_{i,k,j}=h_{j,i,k} \text{ and }
  h_{i,i,j}=0.
\end{equation}
Here, the symmetry requirement leaves ten independent parameters with
$i\le j \le k$, so that after the three partial trace constraints are
included, seven free parameters are left as expected.

We will describe the desired paths of zero field through the
intersection by two curves $\icurvi{\curvidx}(s)$, $\curvidx=1,2$,
parametrized by path length $s$. The requirements that the curves
intersect at the origin and that $\erf$ vanishes on the paths can then
be expressed as
\begin{subequations}
  \label{eq:task}
  \begin{gather}
    \label{eq:icurv-text}
    \icurvi{\curvidx}(0)=\vec{0},\\
    \label{eq:erfic-textf-all}
    \erf(\icurvi{\curvidx}(s))=\vec{0}, \text{for all $s$ and $\curvidx=1,2$}.
  \end{gather}
\end{subequations}
To establish the resulting constraints on the multipole coefficients of $\erf(\vec{r})$,
we will expand $\erf(\icurvi{\curvidx}(s))$ in $s$ and look at each order in
turn. 
We will assume that the curves have a parametrization with continuous
second-order derivatives. In this case, we can require the curves to
be parametrized by path length, so that $\left\lvert \partial_s
  \icurvi{\curvidx}(s) \right\rvert=1$.
Also, according to Eq.~\eqref{eq:icurv-text} we then have that
\begin{equation}
  \label{eq:icurvexp}
  \icurvi{\curvidx}(s)=\tveci{\curvidx} s + \nveci{\curvidx} \curvi{\curvidx} s^2 + \bigo(s^3),
\end{equation}
where, $\tveci{\curvidx}$ and $\nveci{\curvidx}$ are the tangent and normal vectors
of $\icurvi{\curvidx}$ at the origin, and $\curvi{\curvidx}$ is the curvature at the
same point.
Except for the requirement that the tangent and normal vectors are
perpendicular unit vectors, $\tveci{\curvidx}$, $\nveci{\curvidx}$, and
$\curvi{\curvidx}$ can be chosen freely.

Let us now assume that $\erf(\icurvi{\curvidx}(s))=\vec{0}$ as
required by Eq.~\eqref{eq:erfic-textf-all} and deduce the possible
value of the multipole expansion coefficients appearing in
Eq.~\eqref{eq:erfexp}.
Inserting $\icurvi{\curvidx}$ in the form Eq.~\eqref{eq:icurvexp} into the
expansion of $\erf$ as given by Eq.~\eqref{eq:erfexp}, we find that
\begin{equation}
  \label{eq:zerothorder}
  E_i(\icurvi{\curvidx}(s)) = d_i+\bigo(s).
\end{equation}
For $\erf(\icurvi{\curvidx}(s))$ to vanish to zeroth order in $s$, we must
consequently have that
\begin{equation}
  \label{eq:dconst}
  d_i=0,
\end{equation}
so that the homogeneous field component at the origin vanishes.
Continuing the expansion of $E_i(\icurvi{\curvidx}(s))$, given that the $d_i$
fulfill Eq.~\eqref{eq:dconst}, we find
\begin{equation}
  \label{eq:erffirstorder}
  E_i(\icurvi{\curvidx}(s)) = 
  q_{i,j} \tvecij{\curvidx}{i} s + \bigo(s^2).
\end{equation}
It follows that for $\erf(\icurvi{\curvidx}(s))$ to vanish to first order in
$s$, we must require that $\tveci{\curvidx}$ is an eigenvector of
$q_{i,j}$ with eigenvalue $0$ for $\curvidx=1,2$.
In the case where $\tveci{1}$ and $\tveci{2}$ are linearly independent
(the paths are not cotangential at the intersection), it follows from
the additional constraint that $q_{i,j}$ is symmetric and traceless
[Eq.~\eqref{eq:quadconstr}] that
\begin{equation}
  \label{eq:quadconst}
  q_{i,j}=0.
\end{equation}
A more detailed argument, presented in the Appendix, shows that the
quadrupole components must vanish even in the cotangential case
provided the paths are not identical.
For now, we assume the dipole and quadrupole terms to vanish as required by
Eqs.~\eqref{eq:dconst} and \eqref{eq:quadconst}, so that the 
expansion of $E_i(\icurvi{\curvidx}(s))$ to second order takes the form
\begin{equation}
  \label{eq:erfsecorder}
  E_i(\icurvi{\curvidx}(s)) = 
  \tfrac{1}{2} h_{i,j,k} \tvecij{\curvidx}{j} \tvecij{\curvidx}{k} s^2
  + \bigo(s^3).
\end{equation}
Note that the expansion only depends on the tangent vectors
$\tveci{\curvidx}$ of the curves.
For algebraic simplicity, we will orient the coordinate system so that
\begin{subequations}
  \label{eq:tdef}
  \begin{align}
    \tveci{1} &=\cos(\theta)\, \vhat{x} + \sin(\theta)\, \vhat{y}\\
    \tveci{2} &=\cos(\theta)\, \vhat{x} - \sin(\theta)\, \vhat{y},
  \end{align}
\end{subequations}
where $\theta$ is half the angle between the tangents of the paths at
the origin. Without loss of generality we require $0<\theta \le\pi/4$. 
In this case, we find by Eq.~\eqref{eq:erfsecorder} that
\begin{multline}
  \label{eq:erfdiff}
  E_i\left(\icurvi{1}(s)\right) - E_i\left(\icurvi{2}(s)\right) =\\ 
  2  h_{i,1,2} \cos(\theta) \sin(\theta)
   s^2 +\bigo(s^3), 
\end{multline}
and consequently require $h_{i,1,2}=0$ for $E_i(\icurvi{\curvidx})$ to
vanish to second order. Similarly, we find that
\begin{multline}
  \label{eq:erfsum}
  E_i\left(\icurvi{1}(s)\right) + E_i\left(\icurvi{2}(s)\right) =\\ 
  \left[
    h_{i,1,1} \cos^2 (\theta) + h_{i,2,2} \sin^2 (\theta)
  \right] s^2 +\bigo(s^3),
\end{multline}
so that since by Eqs.~\eqref{eq:hexconstr} and \eqref{eq:erfdiff}
$h_{1,1,2}=h_{1,2,2}=0$ also  $h_{2,2,2}$ and $h_{1,1,1}$ must
vanish. Taking the partial trace constraint Eq.~\eqref{eq:hexconstr}
into account, we find that the only hexapole terms that can be nonzero
are $h_{1,1,3}$, $h_{2,2,3}$, and $h_{3,3,3}$.
For these terms we have, according to Eqs.~\eqref{eq:hexconstr} and
\eqref{eq:erfsum}, that $\erf(\icurvi{\curvidx}(s))$ vanishes to second
order in $s$ if and only if
\begin{subequations}
  \label{eq:finaleq}
  \begin{gather}
    h_{1,1,3} \cos^2(\theta)+ h_{2,2,3} \sin^2(\theta) =0\\
    h_{1,1,3} + h_{2,2,3} +h_{3,3,3} =0, 
  \end{gather}
\end{subequations}
with the unique solution that
\begin{equation}
  \label{eq:coefsol}
  \begin{pmatrix}
    h_{1,1,3}\\h_{2,2,3}\\h_{3,3,3}
  \end{pmatrix}=
  6\,\strengthx  
  \begin{pmatrix}
    -\sin^2(\theta)\\
    +\cos^2(\theta)\\
    \sin^2(\theta)-\cos^2(\theta)
  \end{pmatrix},
\end{equation}
for some constant $\strengthx$.
To fulfill the requirements of
Eqs.~(\ref{eq:dconst}), (\ref{eq:quadconst}), and (\ref{eq:coefsol}), we
find that the electrical potential must be of the form

\begin{equation}
  \label{eq:sneakyphi}
  \ve(\vec{r}) = \strengthx \basisx(\vec{r}) + \bigo(\vec{r}^4),
\end{equation}
where the basis function $\basisx$ is given by
\begin{align}
  \basisx(\vec{r})
  &\equiv
  \sin^2(\theta)\, z (3 x^2-z^2)
  - \cos^2(\theta)\,z  (3 y^2-z^2)\\
  &=\sin^2(\theta)\, \basisxx(\vec{r})
  - \cos^2(\theta)\,\basisxy(\vec{r}),
\end{align}
where we have introduced the basis functions $\basisx\topp{i}\equiv z
(3 r_i^2-z^2)$ for later reference.

To summarize, we have constructively established that
to form a zero-field intersection, $\erf(\vec{r})$ must have vanishing
dipole and quadrupole terms.
Furthermore, if the intersection is not cotangential, the only
possible field configuration is
$\erf(\vec{r})=-\strengthx \vec{\nabla}\basisx(\vec{r}) + \bigo(\vec{r}^3)$.

The key assumption in the arguments presented above is that the curves
$\icurvi{\curvidx}$ have continuous second derivatives, so that the
expansion Eq.~(\ref{eq:icurvexp}) is valid.
We assume it to be the case that all isolated lines of zero field will
have this property, but even though the structure of zero-field points
in free space electrostatic potentials, or equivalently critical
points of harmonic functions, has been intensely studied, most results
appear to pertain to 2D systems and we have not found any direct
argument to support our assumption, which must consequently stand as a
conjecture.
On the other hand, we can say for certain that if the field vanishes
on an interval of a curve with analytic parametrization, e.g., a
straight line segment, then it must vanish everywhere on the
analytic  continuation of that line segment. This follows from the
fact that the maps $s \rightarrow E_i(\icurvi{\curvidx}(s))$ will in
this case be analytic as long as $\icurvi{\curvidx}(s)$ is in free
space.  This clearly rules out the possibility of ideal straight-line
Y or T intersections where a straight line segment with zero field
would have to terminate at the intersection point.

Finally, we should point out that a fully confining, zero-field,
right-angle 
intersection does exist. According to the results above, for
this case $\ve(\vec{r})$ must be $\bigo(\vec{r}^4)$, so that the resulting
intersection has a number of disadvantages compared to hexapole
intersections as discussed in the beginning of this section.
By brute force search in the
possible octupole terms we find that $\ve(\vec{r})=\alpha_{\text{O}}
\Theta_{\text{O}}(\vec{r})$ for 
\begin{equation}
  \label{eq:octox}
  \Theta_{\text{O}}(\vec{r})=
    z^4-3 \left(x^2+ y^2\right) z^2+3 x^2 y^2
\end{equation}
is an example of an octupole potential giving rise to an intersection
with zero field on the $x$ and $y$ axes but nowhere else.

\section{Properties of the hexapole intersection}
\label{sec:discussion}

\begin{figure}
  \centering
  \includegraphics[width=\linewidth]{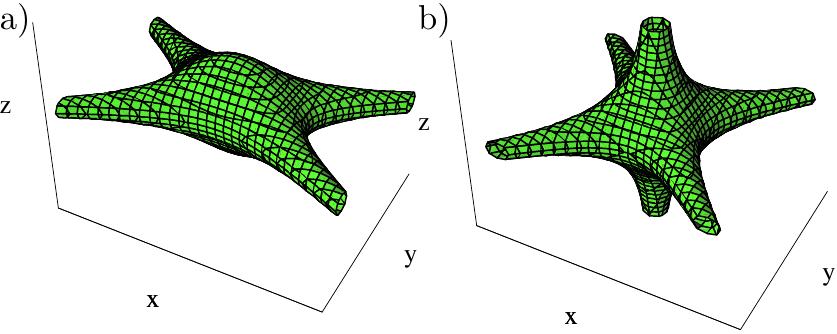}

  \caption{ 
    (Color online) Isosurface of $\left\lvert \grad \basisx
    \right\rvert^2$ for $\basisx$ of Eq.~\eqref{eq:sneakyphi}
    for $\theta=\pi/6$ (a) and $\pi/4$ (b).  
    The isosurface is truncated to a cube: according to
    Eq.~\eqref{eq:tanglines} the zero-field lines extend to infinity.
    Note that there is no confinement along the $z$ axis for
    $\theta=\pi/4$.  
    Since all terms of $\basisx$ are third order in the coordinates, the
    isosurfaces of $\left\lvert \grad \basisx \right\rvert^2$ are related
    by scaling about the origin.  
  }
  \label{fig:isosurf}
\end{figure}

In this section we investigate the properties of the hexapole
intersection described by Eq.~\eqref{eq:sneakyphi} in more detail,
first with respect to confinement properties and then with respect to
implementation considerations. 

\subsection{Confinement properties}
\label{sec:conf-prop}

To study the confinement properties of $\vpp\propto \lvert \grad \basisx\rvert^2$ corresponding to the
hexapole field described by $\basisx$, we look for points where the
corresponding field
\begin{equation}
  \label{eq:gradvex}
  \grad \basisx(\vec{r}) = 3 
  \begin{pmatrix}
    2 x z\, \sin^2(\theta)\\
    -2 y z\, \cos^2(\theta)\\
    (z^2-y^2) \cos^2(\theta) -  (z^2-x^2) \sin^2(\theta)
  \end{pmatrix}
\end{equation}
vanishes.
In the general case, this is seen to happen along the two lines
\begin{equation}
  \label{eq:tanglines}
  y=\pm \tan(\theta) x\quad \text{for $z=0$},
\end{equation}
which are exactly the tangent lines specified in Eq.~\eqref{eq:tdef}.
However, for the special case of $\theta=\pi/4$, $\grad \basisx$ also
vanishes everywhere on the $z$ axis, as illustrated in
Fig.~\ref{fig:isosurf}.

To understand the transition to vanishing confinement in the $z$
direction, we consider the strength of the (quartic) confinement
provided by $\vpp$ along the $y$ and $z$ axes as follows:
\begin{equation}
  \label{eq:quarticstrength}
  \left\lvert \grad \basisx \right\vert^2 = 
  9 
  \begin{cases}
    y^4 \cos^4(\theta)&\text{for $x=z=0$}\\
    z^4 \cos^2(2 \theta)&\text{for $x=y=0$.}
  \end{cases}
\end{equation}
Although confinement strength decreases monotonously with
increasing $\theta$ for both axes, only the $z$ confinement vanishes
completely.

\subsection{Implementation considerations}
\label{sec:impl-cons}
One reason that low-order intersections are interesting is that in
practical implementations, it is difficult to passively ensure that
all lower-order terms vanish exactly.  For typical traps with
nonadjustable electrode geometry and a single rf feed only the
strength and frequency of the rf field can be changed once the trap is
in its operational configuration. The geometry of the rf field,
including the relative strength of the multipole terms at the
intersection point, is not adjustable for such traps.
Since lower-order terms dominate at the intersection point, this could
be critical to the performance of the intersection.
We will consequently investigate which lower-order terms are allowed by
symmetry and try to describe the perturbing effects of these on the
hexapole intersection, Eq.~\eqref{eq:sneakyphi}.

Let us first note that the hexapole intersection potential described
by $\basisx(\vec{r})$ is symmetric in $x$ and $y$ and antisymmetric in $z$,
\begin{subequations}
    \label{eq:fullsym}
  \begin{gather}
    \basisx(x,y,z)=
    \basisx(-x,y,z)=
    \basisx(x,-y,z),\\
    \basisx(x,y,z)=  -\basisx(x,y,-z).
  \end{gather}
\end{subequations}
An analysis similar to that of Sec.~\ref{sec:uniq-hexap-inters} shows
that these symmetry requirements alone constrain $\ve$ to be of the
form
\begin{equation}
  \label{eq:sneakyerr}
  \ve(\vec{r}) =\strengthhomo\,  z +\strengthx\, \basisx(\vec{r})+ \bigo(\vec{r}^4)
\end{equation}
for some constant $\strengthhomo$.
Under the generous assumption that an implementation exactly obeys the
symmetries, Eq.~\eqref{eq:fullsym}, the only possible deviation from
the ideal intersection given by Eq.~\eqref{eq:sneakyphi} is
consequently a homogeneous field in the $z$ direction.
To empirically design a zero-field intersection under this symmetry,
it is thus sufficient to ensure that the field vanishes at the
intersection point, so that $\strengthhomo=0$. 

\begin{figure}
  \includegraphics[width=\linewidth]{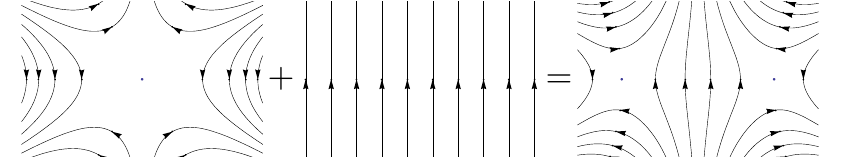}
  \caption{ Field lines of a hexapole and a homogeneous field,
    together with the field lines of the superposition field which has
    two quadrupole poles. Reversing the sign of the homogeneous field
    would shift the two quadrupoles to above and below the hexapole
    center.}
  \label{fig:hexapolesplit}
\end{figure}

In realistic implementations, any field component not explicitly
disallowed by symmetry would most likely be nonzero. It is
consequently of interest to investigate the properties of $\vpp$
corresponding to the form of $\ve$ given by Eq.~(\ref{eq:sneakyerr})
for small nonzero values of $\strengthhomo$.
An intuitive understanding of the effects of a perturbing
homogeneous field can be had by considering the form of $\grad
\basisx$ 
in the $y$-$z$ plane. Here by Eq.~\eqref{eq:gradvex},
\begin{equation}
  \grad \basisx(\vec{r}) = 3
  \begin{pmatrix}
    0\\
    - 2 y z \cos^2(\theta)\\
    z^2 \cos(2 \theta) - y^2 \cos^2(\theta)
  \end{pmatrix} \text{for $x=0$,}
\end{equation}
describing a deformed in-plane hexapole field.
As illustrated by Fig.~\ref{fig:hexapolesplit}, a homogeneous field
will split a hexapole field into two quadrupoles. For a field of the
form given by Eq.~(\ref{eq:sneakyerr}), the positions of these
quadrupole field zeros are
\begin{equation}
  \label{eq:doublejunction}
   \sqrt{\frac{\lvert \strengthhomo \rvert }{3 \lvert \strengthx \rvert}}
  \begin{cases}
    \pm \sec(\theta)\, \vhat{y} &\text{for $\strengthhomo/\strengthx\ge0$}\\
    \pm \sqrt{\sec(2 \theta)}\,\vhat{z} &\text{for $\strengthhomo/\strengthx\le0$.}
  \end{cases}
\end{equation}
The positions are seen to depend critically on the sign of
$\strengthhomo$. For $\strengthhomo/\strengthx > 0$, the zeros are
located on the $y$-axis, while for $\strengthhomo/\strengthx < 0$ they
are located on the $z$-axis.

\begin{figure}
  \includegraphics[width=\linewidth]{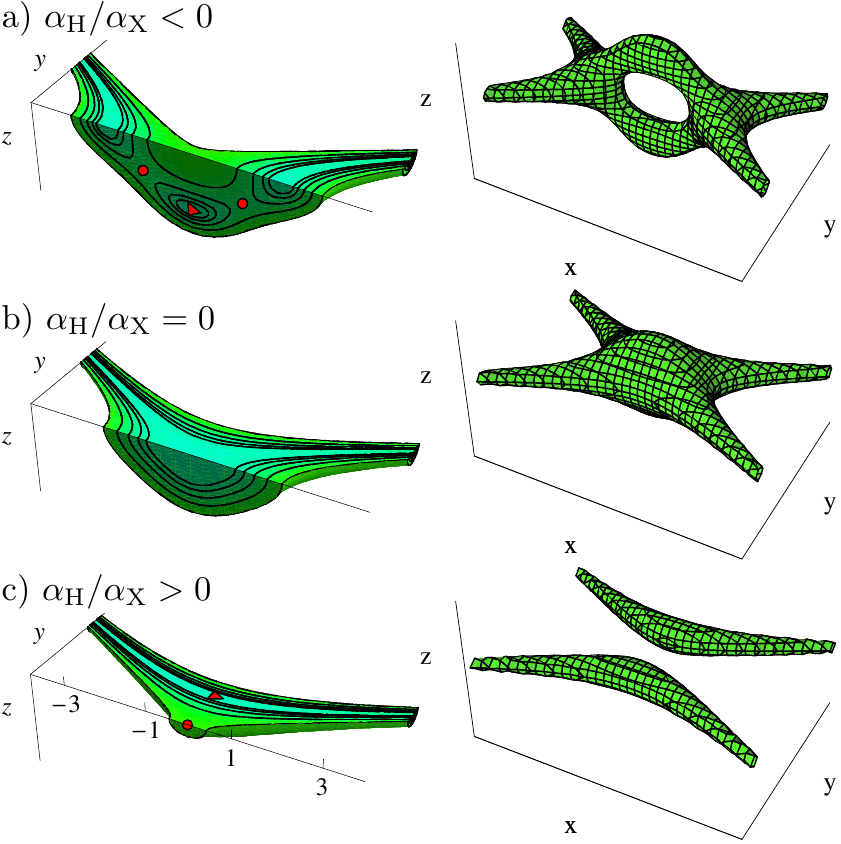}

  \caption{ %
    (Color online) %
    Intersection structure modifications in the presence of a homogeneous
    field component, as described by $\strengthhomo \neq 0$ in
    Eq.~\eqref{eq:sneakyerr}.  
    Left-hand column: Contour curves of $\lvert \grad
    (\strengthhomo z + \strengthx \basisx) \rvert^2$ on $x$-$y$ and
    $x$-$z$ cut planes. Circles mark the barriers with a position given by
    Eq.~(\ref{eq:doublejuntionbumps}) for $\strengthhomo/\strengthx<0$,
    triangles mark the field zeros in the $y$-$z$ plane as described
    by Eq.~(\ref{eq:doublejunction}).
    The length scale is $\sqrt{\lvert \strengthhomo/3 \strengthx
      \rvert}$ and $\theta=\pi/6$.   
    Right-hand column: Full 3D form of the
    next-to-outermost contour.
    Note that in all three cases the outermost contour is connected
    and confined transport between any of the intersection entries is
    possible although the height of the barriers (circles)
    differs as discussed in the text.  }
  \label{fig:snapper}
\end{figure}

The global shape of the intersection described by
$\ve(\vec{r})=\strengthhomo z+\strengthx \basisx(\vec{r})$ 
is very
different in these two cases as illustrated by Fig.~\ref{fig:snapper}.
For $\strengthhomo/\strengthx<0$, we get a double-junction structure with two junctions
on the $x$ axis, connected by paths in the $x$-$z$ plane. The
field is zero on the paths in the $x$-$y$ plane, but this
is not the case for the paths joining the junctions where we find
barriers with a maximum height of
\begin{equation}
  \label{eq:doublejunctionheight}
  \lvert \grad \ve \rvert^2 = 
  \strengthhomo^2 \tan^2(\theta),
\end{equation}
located at the four points (all sign combinations)
\begin{equation}
  \label{eq:doublejuntionbumps}
  \sqrt{\frac{\lvert \strengthhomo \rvert }{6 \lvert \strengthx
      \rvert}}
  \,
  \left(
    \pm \sqrt{\csc^2(\theta)-\sec^2(\theta)}\,\vhat{x}
    \pm \sec(\theta)\,\vhat{z}
  \right).
\end{equation}
This structure is similar to that described in
Refs.~\cite{cassettari00:beam_split_for_guided_atoms,muller00:Waveguide}.
For $\strengthhomo/\strengthx>0$, we get two disjoint paths through the
intersection. The height of the barrier at the origin is
\begin{equation}
  \label{eq:disjointjunctionheight}
  \lvert \grad \ve \rvert^2 = 
  \strengthhomo^2,
\end{equation}
corresponding to the ponderomotive potential due to the homogeneous
field component.

Given a certain production tolerance,
Eqs.~\eqref{eq:doublejunctionheight} and
\eqref{eq:disjointjunctionheight} show that it is beneficial to aim
for a negative value of $\strengthhomo/\strengthx$, corresponding to the double-junction
configuration. For a given value of $\lvert\strengthhomo\rvert$, this will
reduce the barrier height by a factor of $\tan^2(\theta)$ compared to
the case of $\strengthhomo/\strengthx>0$.

\section{Implementation with 3D electrode structure}
\label{sec:impl-with-3d}

\begin{figure}
  \includegraphics[width=\linewidth]{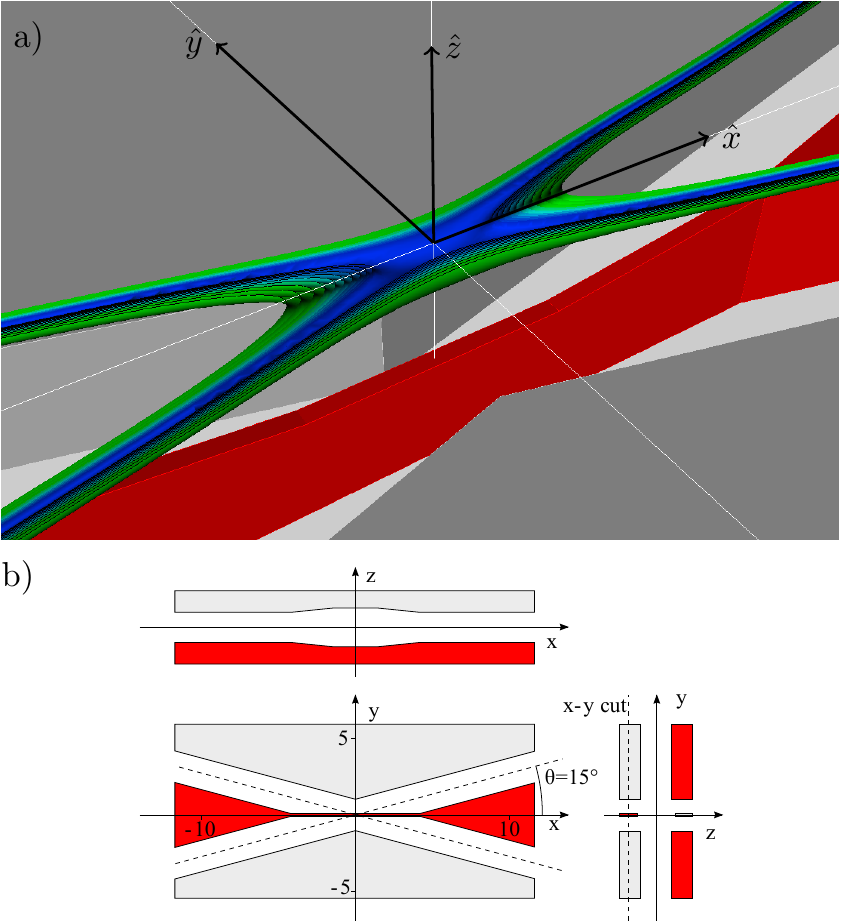}%
  \caption{
    \label{fig:implementation_3D}
    (Color online) Example implementation of an ideal intersection
    using a 3D electrode structure. 
    (a) 3D view of the electrodes (red and gray polygons) and isosurfaces of
    $\lvert\grad \ve(\vec{r}) \rvert^2$ (blue to green curved surfaces) with all
    components with $z>0$ cut away.
    (b) Cuts through the electrode structure, which is symmetric under
    inversion in the $x$-$z$ and $y$-$z$ planes, and antisymmetric
    under inversion in the $x$-$y$ to maintain the symmetry of
    $\basisx$ as described by Eq.~(\ref{eq:fullsym}). Units are the
    ion-electrode distance $d$.
    Results are based on numerical simulations performed using the CPO
    program \cite{chargedparticleoptics}.
  }
\end{figure}

We will now describe a 3D electrode structure that can be used to
implement the ideal intersection described by
Eq.~(\ref{eq:sneakyphi}).

Firstly, we note that the full symmetry of $\basisx$ as described by
Eq.~\eqref{eq:fullsym} can be retained in an implementation based on a
3D configuration of electrodes as illustrated in
Fig.~\ref{fig:implementation_3D}.  According to the results of the
previous section, it follows that the potential at the origin must be
of the form given by Eq.~\eqref{eq:sneakyerr}.
In addition, it turns out, the symmetry ensures unbroken lines of
zero field in the $x$-$y$ plane. By Eq.~\eqref{eq:fullsym} the field
in the $x$-$y$ plane will be along the $z$ direction since the $x$ and
$y$ components of the field will be odd in $z$. Furthermore, the $z$
component of the field in the $x$-$y$ plane is a continuous function
of $x$ and $y$ and the zero contour of this function, consisting of
unbroken lines, will define the points of zero field.

In total, these properties ensure that it is relatively easy to design
a zero-field intersection using 3D electrodes: After choosing an
overall design that ensures that there are paths of zero field leaving
the intersection point, all that remains is to ensure that
$\strengthhomo \equiv -E_z(\vec{0})=0$.
In the implementation illustrated in Fig.~\ref{fig:implementation_3D},
this was achieved by a bisecting search for the correct value of a
single dimensional parameter.

It should be noted that symmetry alone does not define the angle
$\theta$ between the intersecting paths, but rather constrains the
hexapole component to be a superposition of two components
$\basisxx(\vec{r})$ and $\basisxy(\vec{r})$ as follows:
\begin{equation}
  \label{eq:1}
  \ve(\vec{r})=\strengthxx \basisxx(\vec{r})+\strengthxx \basisxx(\vec{r}).
\end{equation}
For the implementation illustrated in
Fig.~\ref{fig:implementation_3D}, we numerically find that
$\strengthxx=0.011\, V_0/d^3$ and $\strengthxy = 0.12 \, V_0/d^3$,
where $V_0$ is the peak amplitude of the applied rf and $d$ is the
distance from the trap center to the nearest electrode.
These values correspond to the rf field being described by
Eq.~(\ref{eq:sneakyphi}) with a a strength of $\strengthx=0.13\,
V_0/d^3$ and an angle of
$\theta=\arctan(\sqrt{\strengthxx/\strengthxy})=17^\circ$.
The obtained strength seems reasonable, given that the strongest
possible hexapole guides along the $x$ or $y$ axes would correspond to
$\strengthx\topp{x/y}=\tfrac{1}{2} V_0/d^3$
\cite{wesenberg07:analytical_theory_multipole}.
The angle is relatively close to the asymptotic angle of
$\theta=15^\circ$ chosen for the demonstration design. If better
agreement was desired, an additional fit parameter (for example, the
length of the bridge recession) could have been included to ensure
that the unwanted hexapole component 
$\cos^2(\theta) \Theta_X\topp{x}+\sin^2(\theta) \Theta_X\topp{y}$ associated with a
given value of $\theta$ was made to vanish.

\section{Conclusions and outlook}
\label{sec:conclusion}

In conclusion, we have shown that a zero-field intersection for rf
traps cannot have any field component of quadrupole or lower order at
the intersection point.
Furthermore, we have demonstrated that the hexapole field component at
the intersection point is uniquely determined by the intersection
angle [Eq.~\eqref{eq:sneakyphi}], and that this component does not
provide any confinement perpendicular to the intersecting paths if
these intersect at right angles.
These results can serve as a guide for the design of intersections for
rf trap networks suitable for QIP based on trapped ions. We have
suggested how the intersection could be implemented using 3D
electrodes.
In relation to implementations of the intersection, we have shown that
if static field components cannot be guaranteed to completely vanish
at the intersection point, it is beneficial to ensure that the
imperfect intersection is of the ``double-junction'' type
[Fig.~\ref{fig:snapper} (a)].

The proposed implementation of the ideal intersection illustrated in
Fig.~\ref{fig:implementation_3D} is not well suited for
microfabrication, which would most likely be a requirement for
fabrication of large scale trap networks. 
For this reason it would be
worthwhile to find intersection implementations with electrode
geometries better suited for microfabrication, such as the
surface-electrode geometry
\cite{chiaverini05:surfac_elect_archit_for_ion}, which is compatible
with large scale microfabrication
\cite{kim05:system_desig_for_large_scale}, and has the lowest
demonstrated heating rates for microfabricated traps 
\cite{seidelin06:microf_surfac_elect_trap_scalab,%
epstein07:simpl_ion_heatin_rate_measur,%
labaziewicz08:suppression}.
Initial investigations in this direction indicate that such
intersections need to be very oblique ($\theta \approx \pi/12$) to
obtain reasonable hexapole strengths for an intersection of two
straight guides. As an alternative, it might be worthwhile to
investigate cotangential intersections of curved guides.

\begin{acknowledgments}
  We thank J. Amini, D. Leibfried, and D. Wineland for stimulating
  discussions.
  This work was supported by the Danish National Research Agency, the Carlsberg
  Foundation, and the QIP IRC (Grant No. GR/S82176/01).
\end{acknowledgments}

\appendix

\section{General proof of absence of quadrupole terms in an intersection}
\label{sec:gener-proof-absence}
\allowdisplaybreaks

Here we present a general proof that there can be no quadrupole terms
in the multipole expansion of $\erf$ as given by
Eq.~\eqref{eq:erfexp}, even in the cotangential case where
$\tveci{1}=\tveci{2}$, provided that the paths $\icurvi{\curvidx}$
have analytic parametrizations and are not identical.

We expand the paths as
\begin{subequations}
  \label{eq:pathexp}
  \begin{align}
    \icurvi{1}(s)&=\sum_{n=1}^\infty \vec{u}\topp{n}\, s^n\\
    \icurvi{2}(s)&=\sum_{n=1}^\infty \vec{v}\topp{n}\, s^n,
  \end{align}
\end{subequations}
where in particular $\vec{u}\topp{1}=\tveci{1}$ and
$\vec{v}\topp{1}=\tveci{2}$ are the identical tangent vectors.
We now take $M$ to be the smallest integer so that $\vec{u}\topp{M}
\neq \vec{v}\topp{M}$.  Such an $M$ must exist since the paths are
assumed not to be identical.
%
Assume as in Sec.~\ref{sec:uniq-hexap-inters} that
$\erf(\icurvi{\curvidx}(s))=\vec{0}$ for all $s$ so that by
Eq.~\eqref{eq:dconst}, $d_i=0$.
We then find that
\begin{multline}
  \label{eq:gendiff}
  E_i(\icurvi{1}(s)) - E_i(\icurvi{2}(s))=\\
  q_{i,j} \left(u_j\topp{M} -v_j\topp{M} \right) s^M +\bigo(s^{M+1}),
\end{multline}
so that for the difference to vanish to $M$-th order, we must have
\begin{equation}
  \label{eq:diffinker}
  \vec{u}\topp{M}-\vec{v}\topp{M} \in \kerq,
\end{equation}
where $\kerq$ is the null space of the $q$ tensor.
As in the noncotangential case, we have from
Eq.~\eqref{eq:erffirstorder} that
\begin{equation}
  \label{eq:tinker}
  \vec{u}\topp{1}=\vec{v}\topp{1} \in \kerq.
\end{equation}

On the other hand, 
letting $\bar{\gamma}(s)=\sum_{n<M} \vec{u}\topp{n} s^n$
we have that
\begin{subequations}
  \label{eq:genspeed}
  \begin{align}
    \left\lvert \partial_s \icurvi{1}(s) \right\rvert^2&=
    \left\lvert \partial_s \bar{\gamma}(s) \right\rvert^2\notag\\
    &+2 \vec{u}\topp{1}\cdot \vec{u}\topp{M} s^{M-1} 
    +\bigo(s^M),
\\
    \left\lvert \partial_s \icurvi{2}(s) \right\rvert^2 &=
    \left\lvert \partial_s \bar{\gamma}(s) \right\rvert^2\notag\\
    &+2 \vec{v}\topp{1}\cdot \vec{v}\topp{M} s^{M-1} 
    +\bigo(s^M).
  \end{align}
\end{subequations}
Taking the difference, we find by Eq.~\eqref{eq:tinker}
\begin{multline}
  \left\lvert \partial_s \icurvi{1}(s) \right\rvert^2
  -
  \left\lvert \partial_s \icurvi{2}(s) \right\rvert^2
  =\\
  2 \vec{u}\topp{1}\cdot \left(\vec{u}\topp{M} - \vec{v}\topp{M}\right) s^{M-1} 
  +\bigo(s^M),  
\end{multline}
so that for the difference to vanish, which it must since both curves
are parametrized by path length, we must have
\begin{equation}
  \label{eq:tvec-leftv-vecvt}
  \vec{u}\topp{1}\cdot \left(\vec{u}\topp{M} - \vec{v}\topp{M}\right)=0, 
\end{equation}
so that $\vec{u}\topp{1}$ and $\vec{u}\topp{M} -
\vec{v}\topp{M}$ are linearly independent. As $\vec{u}\topp{1}$ and
$\vec{u}\topp{M} - \vec{v}\topp{M}$ are both non-zero and by
Eqs.~(\ref{eq:diffinker}) and (\ref{eq:tinker}) members of $\kerq$, it
follows that $q_{i,j}=0$ since $q_{i,j}$ is symmetric and traceless.

\providecommand{\href}[2]{#2}\begingroup\raggedright\endgroup
\end{document}